# Detection of gene communities in multi-networks reveals cancer drivers

Laura Cantini[1,2] *, Enzo Medico[1,3]*, Santo Fortunato[4] and Michele Caselle[5]

[1]Università di Torino, Department of Oncology, Candiolo, Italy

[2]Politecnico di Torino, Department of Control and Computer Engineering, Torino, Italy

[3]Candiolo Cancer Institute, FPO IRCCS, Candiolo, Italy

[4] Department of Computer Science, Aalto University School of Science, Aalto, Finland

[5]Università di Torino, Department of Physics and INFN, Torino, Italy

*Corresponding author:

Laura Cantini, Politecnico di Torino, Department of Control and Computer Engineering, Cso Duca degli Abruzzi 24, 10129 Torino, Italy, Tel. +39-011-0907072  email: lcantini@unito.it

Enzo Medico, Università degli Studi di Torino, S.P. 142, km 3,95 - 10060 Candiolo (TO), Italy

Tel. +39-011-9933234     Fax: +39-011-9933225     email: enzo.medico@unito.it

**We propose a new multi-network-based strategy to integrate different layers of genomic information and use them in a coordinate way to identify driving cancer genes. The multi-networks that we consider combine transcription factor co-targeting, microRNA co-targeting, protein-protein interaction and gene co-expression networks. The rationale behind this choice is that gene co-expression and protein-protein interactions require a tight coregulation of the partners and that such a fine tuned regulation can be obtained only combining both the transcriptional and post-transcriptional layers of regulation. To extract the relevant biological information from the multi-network we studied its partition into communities. To this end we applied a consensus clustering algorithm based on state of art community detection methods. Even if our procedure is valid in principle for any pathology in this work we concentrate on gastric, lung, pancreas and colorectal cancer and identified from the enrichment analysis of the multi-network communities a set of candidate driver cancer genes. Some of them were already known oncogenes while a few are new. The combination of the different layers of information allowed us to extract from the multi-network indications on the regulatory pattern and functional role of both the already known and the new candidate driver genes.**

In the past years the advent of high-throughput experimental technologies provided biologists with a flood of molecular data. This huge amount of information requires the design of efficient methodologies to be interpreted. Among them, network analysis proved to be a very effective approach to capture the molecular complexity of human disease and to discern how such complexity controls disease manifestations, prognosis, and therapy [1]. Thus far, network-based computational methods were primarily focused on the analysis of single biological networks (e.g. protein-protein interaction network, gene co-expression network, and so on). However, the biological relationships described by different networks are in most cases not independent, like in the case of gene co-expression and transcription factor networks. Therefore, studying single networks in isolation turned out to be insufficient to unveil functional regulatory patterns originating from complex interactions across multiple layers of biological relationships. For this reason, a new pressing request in molecular biology is to design network-based methods allowing combined use of multiple levels of genomic and molecular interaction data. Many solutions have been proposed in the last few years (see for instance [2,3]). Among them a special role has been played by multiplex networks which emerged recently as one of the major contemporary topics in network theory [4,5]. A multiplex network is a set of N nodes interacting among them in M different layers, each reflecting a distinct type of interaction linking the same pair of nodes (see Fig. 1). Some relevant applications in biology already exist: Li and colleagues studied a multilayer structure composed of 130 co-expression networks, in which each layer represents a different experimental condition [6]. Subsequently, they also constructed two-layer networks, composed of a standard co-expression network and an exon co-splicing network [7]. More recently, Bennett and co-workers [8] identified communities on the multiplex network of physical, genetic and co-expression interactions, in yeast, using mathematical programming with the modularity by Newman and Girvan as objective function. These studies showed that multiplex networks may be very effective in combining different layers of experimental information. Following this line we propose here a multi-network-based approach for the identification of candidate driving genes in cancer. We use the expression multi-networks instead of multiplex because we will not consider couplings between the layers.

Cancer is a complex disease caused by a progressive accumulation of dysfunctions in neoplastic cells. During the last decade, technological advancements and reducing costs enabled laboratories to quantitatively monitor these alterations. Efficient methodologies were designed to interpret these data and identify the genes driving the neoplastic growth. However these approaches are classically applied to study separately biological measurements that are clearly not independent. For this reason we consider the identification of driver cancer genes as perfectly suited for a multi-network-type analysis. To address this problem, we combined, in a single multi-network, four different gene networks: (i) Transcription Factor (TF) co-targeting network, (ii) microRNA co-targeting network, (iii) Protein-Protein Interaction (PPI) network and (iv) gene co-expression network. The rationale behind this choice is that the insurgence of cancer is typically due to a dysregulation of the signaling and/or of the regulatory network of the cell. These regulatory pathways are tightly controlled in the cell both at the transcriptional and at the post-transcriptional (microRNA) levels[9] and their alteration very often involves modification in the expression levels of genes which are at the same time partners in a protein-protein interaction and targeted by the same set of transcription factors and miRNAs. These are exactly the events which are selected and prioritized in a Multi-network-based analysis like the one that we propose. Following the construction of the multi-network, we proceed with the identification of communities, that is, of groups of nodes that are densely connected to each other, but sparsely connected to the other nodes of the network. This is achieved by detection of gene communities within each multi-network layer and subsequent identification of communities via consensus clustering across the four layers. It is well known that community detection within a network is an open and difficult problem and that different and complementary strategies exist, based on the type of information that one wants to optimize [10]. For this reason we tested some well-known community detection algorithms, Infomap [11], OSLOM [12], Label propagation [13], Louvain [14] and Modularity optimization via simulated annealing [15], all of which can be run in our multi-network analysis package "Gene4x" described in Supplementary Fig. S1 and available at https://github.com/lcan88/Gene4x.git. The package uses as input an expression dataset supplied by the user, creates the four-layer multi-network and provides as output the multi-network community structure. To check whether the multi-network communities are more biologically relevant than the communities obtained in the expression network alone, we applied the analysis to human gastric, lung, pancreas and colon cancer datasets, and tested the resulting multi-network or co-expression network communities for functional enrichment, or differential expression between tumor and normal tissues. In all four cancer types, the multi-network communities were more significant in both tests. As a further check, for each cancer type, we also constructed a multi-network containing as fourth layer the healthy tissue co-expression network, and selected functions enriched in tumor communities but not healthy tissue communities. Again, in all cancer types, the multi-network communities highlighted new relevant tumor-specific functional enrichments, including chromosomal aberrations, candidate markers and driver genes, not detected by the co-expression network alone. Some of the genes that we found were already known oncogenes while a few are new and represent interesting candidate cancer drivers. Moreover, the combination of the different layers of information of the multi-network allowed us to extract useful hints on the regulatory pattern and functional role both for the already known oncogenes and for the new candidate driver genes.

# RESULTS

We organized this section in two parts. The first is devoted to the construction of the different multi-networks and the identification of their community structure. Even if we shall specialize the construction to four specific sets of cancer expression data, the approach is completely general and it could be adapted to any set of expression data. Apart from minor details this first section follows the steps of the freely downloadable pipeline that we provide at https://github.com/lcan88/Gene4x.git. The second section is instead specifically devoted to the results that we obtained on our chosen cancer datasets. We report a set of tests on the obtained communities, supporting our claim that the multi-network structure that we proposed is able to integrate efficiently the information coming from the different layers.

## Construction of the multi-networks and identification of their community structure.

### *Construction of the multi-network.*

The study was conducted separately on four tissues: gastric, lung, pancreatic and colon. For each of them two multi-networks were constructed: one for the normal tissue and one for the tumor. Each multi-network was composed of four layers: co-expression network, transcription factor (TF) co-targeting network, microRNA co-targeting network and protein-protein interaction network (PPI). The nodes of the multi-network are the genes, the interactions in the different layers were obtained as follows:

- The co-expression network was constructed from microarray expression data of the four tissues: gastric [16], lung [17], pancreas [18] and colon [19]. The intra-array normalized expression data were downloaded from GEO database (www.ncbi.nlm.nih.gov/geo/) and quantile normalized. Subsequently probes mapping to the same Entrez gene ID were averaged and finally the matrix was log2-transformed as in [2]. The network reconstruction involved the computation of the mutual information (MI) among all the possible couples of genes, obtaining in this way a complete weighted graph.
- The TF co-targeting network was assembled starting from ENCODE experimentally

validated TF-target interactions (ChIP-seq) [20]. It is a weighted network, with positive integer weights in which a link is introduced between two genes if they share at least one common regulator (TF). The weight of the link is simply the number of TFs targeting both the genes.

- The microRNA's co-targeting network is constructed in a similar way starting from five independent databases of microRNA-target interactions: miRTarBase 2.5 [21], doRiNA-PicTar 2012 [22], microRNA.org 2010 [23], PITA 2007 [24], TargetScan 6.1 [25]. Only those interactions predicted by at least two databases were considered. The reconstruction procedure is the same previously explained for the TF co-targeting network.

- Finally, the protein-protein interaction network (PrePPI) reporting experimentally validated binding between proteins was downloaded from [26], then node names were converted from protein to gene symbol.

We remark that only the layer corresponding to the co-expression network changes in function of the type (gastric/lung/pancreatic/colon) and state (tumor/normal) of the tissue under study, while the other three remain unchanged. The four layers contained different genes thus the last step to obtain a multi-network structure was to extract in each layer the subnetwork composed of only those nodes common to all layers. We obtained in this way a multi-network of around 5,000 nodes for each of the four tissues (5325 in Gastric, 5354 in Lung, 5307 in Pancreas and 5148 in Colon).

*Layers Filtering*

Two of the four layers (TF and microRNA co-targeting networks) had a high link density (around 20% and more than 75% in all the four tissues for the TF and microRNA co-targeting networks, respectively) and one of the four layers (co-expression network) is a complete graph. This is a major obstacle for typical community detection algorithms whose performances are instead optimal on sparse networks. Thus a preliminary mandatory step of the whole analysis was network filtering, in order to decrease the link density of these networks. This filtering step is very delicate, as it must be performed without loosing the biological information contained in the networks. In the field of complex networks, various techniques were proposed to achieve this goal. The simplest choice, which is often used for networks not having a fat-tailed degree distribution, is a global thresholding, that filters links based on the weight distribution. In our multi-network, two of the four layers do not have fat-tailed weight distributions and thus we could in principle use thresholding, which however turned out to be highly ineffective for our networks. As shown in Supplementary Table S1, this filter led to an almost constant high link density (10-30%) even for very stringent values of the threshold. This is due to the particular topology of the co-targeting and co-expression networks in which a filter with a global threshold deletes not only links but also a significant amount of nodes. A much better choice was the disparity filter proposed by Serrano et al. [27]. This filter was originally designed for networks with fat-tailed weight distributions but turned out to be very effective also for our co-targeting and co-expression networks. The disparity filter output depends on the choice of a significance level α that, as suggested by Serrano et al., has to be maintained in the range [0.01,0.5]. The optimal value of α for our networks was chosen following three criteria:

  i. Low density of the output network,
 ii. A balanced number of links among the different layers
iii. The presence of a significant number of validated links among those of the network.

The third criterion was implemented by testing, through a Fisher exact test, the significance of the intersection between the output network and a collection of putative predicted interactions. The predicted interactions were extracted from three main categories of databases:

  a. Interaction databases that include gene/protein interactions validated through biochemical experiments (BioGRID,IntAct),
  b. Pathway database (CELL,REACTOME,IMID)
  c. Databases which contain interactions obtained via a manually curated or a software based mining of the literature (HPRD, MINT, IntAct, ID-serve).

The results of the three criteria for different values of α are reported in Supplementary Table S2, where the optimal α values are highlighted in bold. As for the practical implementation of this choice, it is important to stress that our analysis is rather robust with respect to changes in α and in particular that the results of the community detection are not substantially affected by small changes in α. We tested this by comparing the partition in communities obtained by doubling or halving alpha with respect to the optimal one (see details on the comparison procedure in the Material and Methods section) and in all cases we found an overlap of 99% between the different partitions.

*Community detection in the multi-network*

After filtering, all the layers of the multi-network were sparse enough to perform community detection. The design of community detection algorithms on multi-networks is still an open problem [28]. We propose here a possible solution based on the use of the consensus clustering procedure described in [29]. We used five widely adopted algorithms: Infomap [11], OSLOM [12], Label propagation [13], Louvain [14] and Modularity optimization via simulated annealing [15]. We integrated in our software all five algorithms, leaving the choice of the preferred one to the user. We discuss in the Supplementary Informations a few criteria which can be used to drive this choice, but their use strongly depends on the biological problem at hand and on the size and type of expression data, thus we decided in our pipeline to provide all the five multi-network partitions and leave the final choice to the user. We stress that the algorithms are all stochastic, so they give different partitions for different choices of the random seeds. Therefore, to get the best result for each layer we computed the consensus partition over 100 runs of every algorithm on each layer. Then, we combined the best partitions of the four layers into a single consensus partition, describing the community structure of the multi-network. For all the multi-networks that we studied, the number of communities identified by the different algorithms had very different ranges, from (5-7) for Modularity optimization to (150-170) of OSLOM. The low number of clusters found via modularity optimization is due to the well-known resolution limit of this technique [30]. Consequently, the size of the obtained communities varies as well with the selected detection method (see Supplementary Fig. S2). In particular, clusters found through Modularity optimization via simulated annealing are typically large (> 140 elements). Infomap and Louvain have two peaks in the distribution of cluster sizes, one for small communities (< 10 elements) and one for large communities (> 140 elements). Finally, OSLOM has a more homogeneous community size distribution, with the majority of the communities in the size interval 20-50 elements. The fact that this behavior was reproduced in all the eight multi-networks that we studied indicates that this is most probably a structural feature of the different algorithms and the user should take it into account when choosing one of them for the analysis. In the following we will use in particular the community organization obtained using OSLOM (which we listed in Supplementary Tables S3-S6). These communities (both for the Normal and for the Cancer Multi-network) are the final output of the pipeline that we propose. In the next sections we shall discuss a few relevant features of our results.

**Multi-network vs. single layer communities: structure and biological significance.**

*Multi-network communities have a small overlap with the communities of the individual layers.*

In order to test if the community organization of the multi-network is driven by a single layer or is the result of the combined influence of all the layers we decided, as a first step, to study the overlap between the communities of the multi-network and those extracted (using the same procedure discussed in the previous section) from the single layers. To achieve this goal, we compared the multi-network communities with those of each layer through Normalized Mutual Information (NMI)[31], a measure of correspondence between two partitions. NMI is always in the range [0,1], it equals one for perfect correspondence and zero for no correspondence. As shown in Supplementary Table S7 the NMI of our comparisons is always much lower than one and in some cases it becomes zero. In particular, the multi-network communities have different origins, some of them have a high overlap with one layer and a smaller contribution of the others, while some others are derived from an equivalent mixture of communities contained in all the layers. In general the multi-network communities seem to have a higher resolution with respect to those of the single layers. For instance the number of communities identified in the expression network is always smaller than that in the multi-network. In particular the communities detected on the expression network are: 127 in gastric, 118 in lung, 137 in pancreas and 79 in colon while those identified in the multi-network are: 158 in gastric, 162 in lung, 173 in pancreas and 178 in colon. Taken together these observations support the claim that the community organization of the multi-network is not driven by a single layer but combines information coming from all the layers, in a non-trivial manner.

*Multi-network communities are more informative than those obtained in the expression networks of tumor tissues.*

Among the different layers of the Multi-network a special role is played by the co-expression one since a goal of our analysis was exactly to show that it is possible to gain additional biological information on the selected pathology from the multi-network with respect to that accessible from the expression data alone. To test this claim we thus compared, following the method used to compare the community detection algorithms and outlined in Supplementary Informations, the functional enrichment of the multi-network communities with respect to the co-expression ones. Also in this case we tested only tumor multi-networks for simplicity. In all cases we found that the amount of functionally homogeneous communities in the expression network was smaller than that in the multi-network. In detail, the percentage of functionally homogeneous communities in expression network vs multi-network is the following: gastric 57% vs 63%, lung 71% vs 77%, pancreas 54% vs 67% and colon 86% vs 88%. Finally, we also compared the two community organizations using the second criterion discussed in the Supplementary Informations for community detection algorithms comparison, i.e. checking which of the two types of partition was able to group together in the same community genes differentially expressed comparing tumor and normal tissue. As shown in Fig. 2 in all four tissues the multi-network communities performed better than the co-expression communities. In particular, in gastric and pancreas the multi-network performs better than the expression network in all the differential expression tests.

*Multi-network communities are enriched in biological components involved in the oncogenic process that one could not get from the expression networks alone.*

At the end, we compared the performances of the two networks in the identification of communities involved in the oncogenic process, a problem of medical interest. The community enrichment procedure that we used is the one described for criterion (i) used above to choose the optimal community detection algorithm, the percentage of functionally homogeneous communities identified. In each of the four tissues, the procedure was applied for all the communities detected in the tumor multi-network (TM), normal multi-network (NM), tumor expression network (TE) and normal expression network (NE). Then we selected, among all the biological functions enriched in the communities of both the multi-network and the expression network, only those characteristic of the neoplastic tissues. This selection was performed subtracting from all the biological categories significant in the communities of the TM those that were also significant in the communities of the NM. The same procedure was followed also for the communities of the expression network. Finally we compared how the communities identified in two networks were able to detect functions associated to the tumor and not to the normal tissue. This was performed considering, among the biological functions retained at the previous step, only those of the multi-network that were not present in the expression network. Moreover, we divided for sake of clarity the biological functions, selected according to the procedure described above, in four categories: chromosomal locations, pathways, motives TF/microRNAs and Gene Ontology (GO). We report in Supplementary Table S8-S12 the results of this analysis. Then, for each category, the significant functions with a p-value lower than $10^{-5}$ are summarized in a radar plot, reported in Figure 3. In general most of these highly enriched categories are tissue specific. The only exception (which we shall discuss in detail below) are the chromosomal location (see Fig.3), which are enriched in more than one type of cancer. We consider the list of enriched categories reported in Supplementary Tables S8-S12 as one of the most important results of our analysis and they will be the starting point of the discussion on the biological implication of our results of the next section.

## DISCUSSION

The main goal of our work was to show the power of multi-network-based methods to identify candidate driving genes in cancer. Among the four cancer types to which we applied our algorithm we discuss here for reason of space only the Pancreatic cancer case, but similar results can be obtained also for the other cancer types. Among the various results obtained with the multi-network analysis we concentrate in particular on three observables which are best suited to identify candidate driving cancer genes: (i) Enriched chromosomal locations, (ii) Intersection between our communities and known cancer signatures and (iii) Enriched miRNA regulons.

Chromosomal location (see Supplementary Table S8) is the class of enriched categories in which the role of the multi-network can be better appreciated. At a first glance one could consider this enrichment as a trivial consequence of the fact that cancer expression data are biased by chromosomal aberrations but this is certainly not our case since the multi-network communities that we are studying are exactly those that cannot be obtained from the expression networks only. This suggests that these chromosomal locations are free of the noise due to random chromosomal aberration of typical expression data and capture the real causative aberrations that are not only present in the expression data but also appear in a consistent way in the regulatory components of the multi-network. Indeed a careful inspection of these enriched chromosomal locations shows that in several cases they correspond to loci that are already known to be associated with the corresponding cancer types. This strongly supports a similar association also for the remaining loci. In this sense multi-network analysis could help to identify still unknown players in the oncogenic process. Moreover, the study of multi-network based enriched chromosomal loci may have three further important outcomes: (i) out of the hundreds of genes contained in each enriched chromosomal location with our analysis we select only the few which are involved in a common co-regulatory scheme and thus are likely to be the real drivers of the cancer; (ii) In the communities we find also genes outside the enriched chromosomal locus, related to them non only by a co-expression link but also by regulatory relations and this suggests that they could be part of a common biological pathway which is dysregulated in the tumour; (iii) In some cases the community is also characterized by a GO or KEGG enriched category which may give some hint to identify the above pathways. These three steps are discussed in detail for the pancreatic cancer dataset in the Supplementary Informations. We report here only a couple of examples of the results. As an example of point (i) the 1q21 locus has an intersection with the 106th community which is composed by five genes: "F11R", "HDGF", "ILF2", "PRCC", "VPS72". Among them, F11R (JAM-A) is associated with metastasis and poor survival in pancreatic cancer [32], HDGF is a prognostic factor for patients with pancreatic cancer [33] and PRCC has been recently shown to be mutated in the pancreatic tumour [34]. It is interesting to note that out of the hundreds of genes contained in this locus we were able to single out three known oncogenes, moreover, it is probably the fact that these oncogenes are located in the same chromosomal locus which makes alterations of this locus so dangerous. As an example of point (iii) the 43th community of the Pancreas dataset has 45 genes, five of which belong to the locus 1q32: ATF3, BTG2, CD46, IRF6 and PPP1R15B. As in the previous cases, also for this locus three out of these five genes ATF3 BTG2 and CD46 are already known markers of pancreatic cancer[35-36]. What is more interesting for our purposes is that in this case we have some more information on the possible pathways in which these genes, and the other belonging to the community, are involved. Looking at the enrichment analysis for this community we find the DREAM pathway which involves the JUN and FOS regulators. Indeed looking at the other genes belonging to the community we find several genes of the JUN and FOS families. All these findings suggest a cooperative role of several genes of the community (not only those belonging to the selected locus but also the other) in the *apoptotic* process and more generally in *cell survival*. Given that the biological details about chromosomal locations and microRNAs found significant in Pancreatic communities need an extended description we reported them in the Supplementary Informations.

Another important application of our analysis is in connection to cancer gene signatures. There are several examples of gene sets proposed as signature of poor/good prognosis or able to distinguish between different grades or types of the cancer under study. If one of these signatures has a significant overlap with one (or more) of our communities then this can be used to enrich the signature and/or to understand the molecular mechanism behind the signature properties (good/poor prognosis or tumor classification). In the pancreatic tumor which we are using as example there is a well established signature [37] which is used to identify different subtypes of the pancreatic ductal adenocarcinoma and to predict their response to therapy. This signature divides the tumor in three subtypes defined as "Classical", "Quasi-Mesemchymal" (QM) and "Exocrine-like" (E). Intersecting these three gene sets with our communities we found that a few communities were remarkably enriched in signature genes (see Table 1). In particular the exocrine-like signature is enriched in the second community which in turn (see Supplementary Table S10) is enriched in targets of the HNF1 Transcription Factor. The importance of HNF1 in pancreatic cancer is known since long time (see for instance[38,39]). What is more interesting for our purposes is that recently a set of genome-wide analyses identified precisely the HNF1 homeobox A (HNF1A) as an essential component of the secretory pathway in the exocrine pancreas. (see for instance[40-42]). Our analysis thus suggests that the exocrine-like signature of Collison et al. could be indeed related to the HNF1 regulome. This suggests a way to enlarge this signature and, as mentioned above, to understand the regulatory pathway behind the signature.

Finally a special role in our list of candidate driving genes is played by miRNAs. It is well known that miRNAs play a crucial role in cancer development [43]. Our multi-network analysis is perfectly suited to identify these potential miRNA drivers since one of the layers is exactly the miRNA cotargeting network and in this case the role of the other layers is to filter among all the miRNAs only those whose targets also interact in one or more of the other layers and in particular in the cancer coexpression one. As a final result (see Supplementary Table S9) we find for each of the cancer types a list of miRNAs enriched in one or more of our communities. We included in the Supplementary Informations a detailed discussion of all these candidates. Some of them are already well known but a few are new and represent one of the most relevant biological outcomes of our analysis. We report here only two examples which we find of particular interest and give an idea of the power of this type of analysis. MiR-337 is enriched with rather strong p-values in three communities. It is a known regulator of HOXB7 and its overexpression induces a significant suppression of Pancreatic Ductal Adenocarcinoma (PDAC) cell proliferation and invasion [44] and in fact it is associated with longer survival in pancreatic cancer [45]. MiR-153 is enriched in the 41$^{st}$ community and is known to inhibit PDAC cell migration and invasion by targeting SNAI1. Also this miRNA is an independent prognostic marker for predicting 3-year survival of PDAC patients[46]. Its role in tumorigenesis was highlighted also through an independent bioinformatics analysis in [47]. These results rise the hope that a similar role as prognostic markers could be played also by the other miRNAs that we found enriched and indeed in some cases we found strong evidences in this direction (see detailed discussion in S.I.).

These results and those on the comparison analysis prove that the use of a multi-network combining Transcription Factor (TF) targeting, microRNA targeting, Protein-protein interaction and gene co-expression across samples gives a significantly richer information than the single expression networks. Altogether our results provide encouraging evidence that multilayer networks contain new information on the structure and dynamics of complex systems, that would not be possible to get from the individual components. Moreover our algorithm is such that it could easily accommodate other sources of genomic data, as epigenetic markers. This, and the recent developments in multilayer network technology [4,5] are likely to improve in the near future our ability to identify driving cancer genes and possibly to extend our methods also to other pathologies.

# Acknowledgements

S. F. acknowledges MULTIPLEX, grant number 317532 of the European Commission

M.C. acknowledges the grant from Compagnia San Paolo/ Progetto di ricerca di Ateneo GeneRNet

E.M. acknowledges the grants from AIRC (IG n. 12944 and 2010 Special Program Molecular Clinical Oncology 5x1000 project n. 9970), and Fondazione Piemontese per la Ricerca sul Cancro-ONLUS (5x1000 Ministero della Salute 2010-OGC and 2011-Implementing genomic-driven precision oncology at the IRCC).

L.C. acknowledges Santo Fortunato and BECS department for the hospitality during the realization of this work. She also acknowledges Santo Fortunato and his research group, in particular Darko Hric, for the methodological support.

## Contributions

L. C. S. F. E. M and M. C have designed the research and written the paper. L. C. has performed the research

## Competing financial interests

The authors declare no competing financial interests.

## Figure Legends

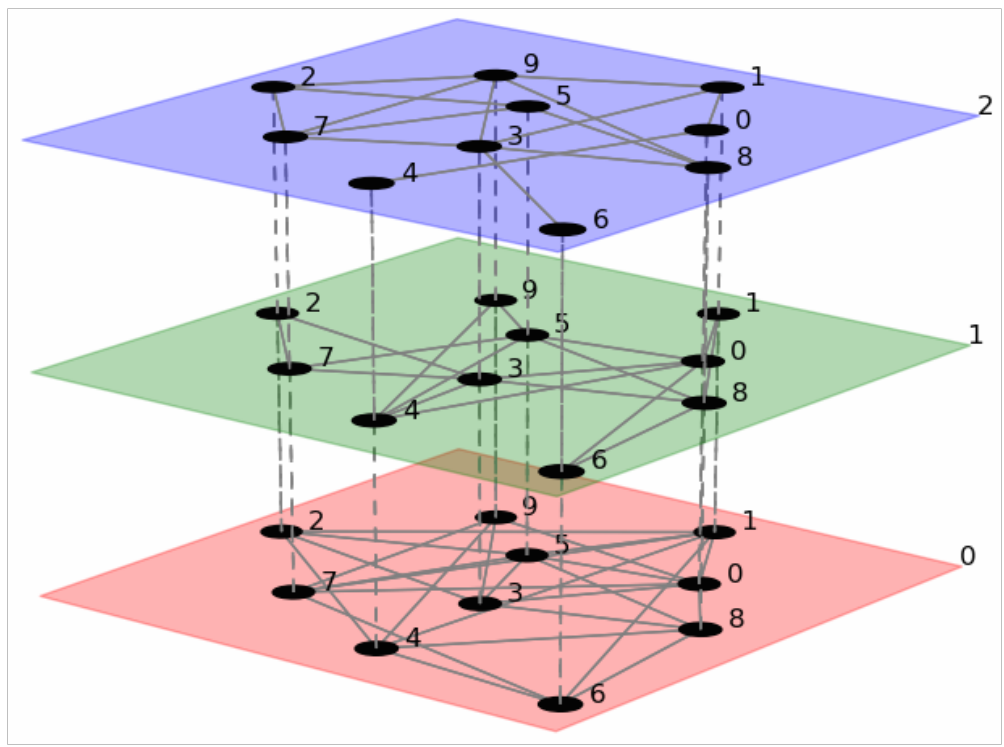

**Figure 1. Example of multi-network.**

Example of multi-network with M=3 layers (represented in red, green and blue) and N=10 nodes. Nodes are the same in all the three layers. Intra-layers links are represented with solid lines, while inter-layer interactions (dashed lines) are from each node to itself in the other layers. The figure was downloaded from http://people.maths.ox.ac.uk/kivela/mln_library/visualizing.html.

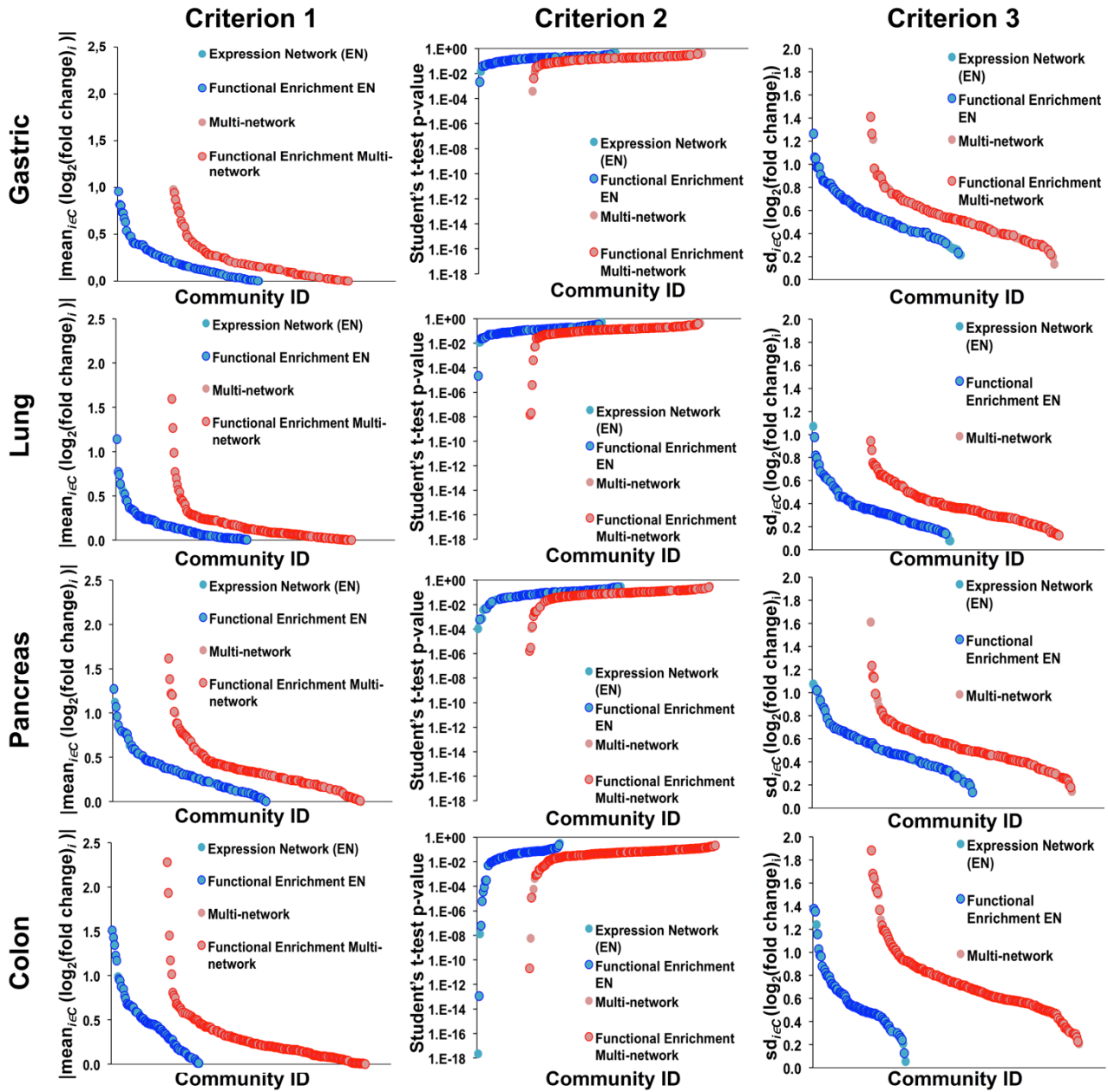

**Figure 2. Comparison between multi-network and expression networks ability to reveal (normal vs. tumor) differentially expressed communities.**

Multi-network (blue) and expression (red) networks were tested in their ability to reveal differentially expressed communities in the comparison between tumor and normal tissue. Each dot in the plot represents a community, a darker color identifies those communities that are also functionally homogeneous. In the columns we report the results of the three differential expression criteria: $|\text{mean}_{i \in C} (\log_2(\text{fold change})_i)|$ (Criterion 1); Student's t-test p-value (Criterion 2); $\text{sd}_{i \in C} (\log_2(\text{fold change})_i)$ (Criterion 3).

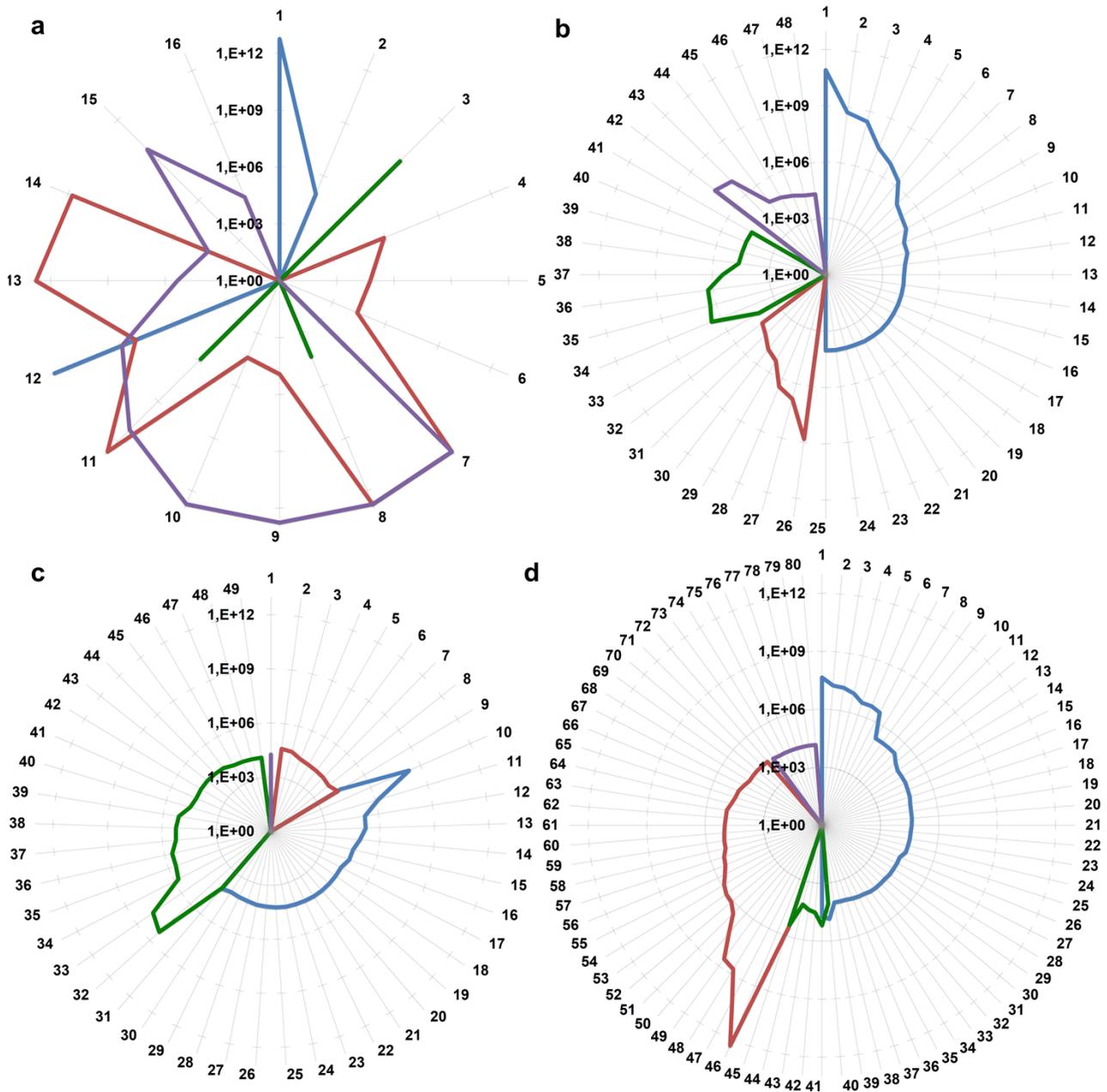

**Figure 3. Biological components involved in the oncogenic process enriched in the multi-network communities that we could not get from the expression network**. Radar plots of the reciprocal of the enrichment p-values for (a) chromosomes, (b) pathways, (c) TF/microRNAs motifs and (d) GO. In each radar plot the results of the enrichment analysis are represented for the four tissues: gastric (blue), lung (red), pancreas (green) and colon (violet). Only the functions with an enrichment p-value lower than 10^-5 are represented. The conversion can be found in Supplementary Table S13.

| class | community ID | p-value | signature size | community size | intersection | genes | | | |
|---|---|---|---|---|---|---|---|---|---|
| exocrine-like | 2 | 9.97E-04 | 2 | 119 | 2 | SLC4A4 | SPINK1 | | |
| QM-PDA | 7 | 3.22E-02 | 12 | 88 | 2 | PHLDA1 | HK2 | | |
| classical | 19 | 8.67E-06 | 10 | 66 | 4 | TFF1 | TFF3 | ERBB3 | GPX2 |

| | | | | | | | |
|---|---|---|---|---|---|---|---|
| classical | 37 | 3.39E-04 | 10 | 49 | 1 | FERMT1 | |
| QM-PDA | 57 | 6.97E-03 | 12 | 40 | 2 | CA | TWIST1 |
| QM-PDA | 162 | 2.70E-02 | 12 | 6 | 1 | S100A2 | |

**The Table reports the intersection of pancreatic communities with Collisson signature. In particular, in the first column is reported the name of gene set within the Collisson signature, in the second the ID of the community that has a significant intersection with the selected Collison gene set, in the third the Benjamini-Hochberg corrected hypergeometric p-value of the intersection, in the fourth the size of the gene set, in the fifth the community size, in the seventh the intersection size, in the eighth the list of genes belonging to the intersection.**